\documentclass[12pt]{article}

\def\pt{p_T}
\def\dis{distribution}
\def\bq{\begin{eqnarray}}
\def\eq{\end{eqnarray}}
\usepackage{graphicx}

\begin{document}

\begin{center}  {\Large {\bf Phenomenological Relationship between the Ridge and Inclusive Distributions}}

\vskip .75cm
 {\bf Charles B.\ Chiu$^1$ and  Rudolph C. Hwa$^{2}$}

\vskip.5cm
{$^1$Center for Particles and Fields and Department of Physics,\\
University of Texas at Austin, Austin, TX 78712, USA\\
$^2$Institute of Theoretical Science and Department of
Physics,\\ University of Oregon, Eugene, OR 97403-5203, USA}
\end{center}

\vskip.5cm

\begin{abstract} 
 A relationship between the ridge \dis\ in $\Delta\eta$ and the single-particle \dis\ in $\eta$ is proposed. It is then verified by use of the data from PHOBOS on both \dis s. The implication seems to point to the possibility that there is no long-range longitudinal correlation. An interpretation of the phenomenological observation along that possibility is developed.
 
\end{abstract}
 
\vskip.5cm
\section{Introduction}
The ridge structure in two-particle correlation has been studied in nuclear collisions at the Relativistic Heavy-Ion Collider (RHIC) for several years \cite{ja, bia, ba, bia1, md} and is now also seen in $pp$ collisions at the Large Hadron Collider (LHC) \cite{cms}.  The nature of that structure is that it is narrow in $\Delta\phi$ (azimuthal angle $\phi$ relative to that of the trigger) but broad in $\Delta\eta$ (pseudorapidity $\eta$ relative to the trigger).  In Ref.\ \cite{bia1} the range in $\Delta\eta$ is found to be as large as 4.  So far there is no consensus on the origin of the ridge formation.  It has been pointed out that the wide $\Delta\eta$ distribution implies long-range correlation \cite{ad,gmm}.  That is a view based mainly on theoretical ideas without any comparison between the $\eta$ ranges of single-particle distribution and two-particle correlation.  We make that comparison here using only experimental data from PHOBOS \cite{ba,bb}. It is found that the large $\Delta\eta$ ridge \dis\ is related simply to a shift of the pseudorapidity \dis\ and an integral over the trigger $\eta$. That is a phenomenological observation without any theoretical input. Any successful model of ridge formation should be able to explain that relationship. 

There are subtleties about the single-particle \dis\ for all charges, $dN^{ch}/d\eta$, that to our knowledge has not been satisfactorily explained in all its details. Since it sums over all charges, hadrons of different types are included, making $dN^{ch}/d\eta$ to be quite different from $dN^{\pi}/dy$. That difference cannot be readily accounted for in any simple hadronization scheme. Fortunately, detailed examination of $dN^{ch}/d\eta$ is not required before we find its relationship to the ridge \dis\ $dN^{ch}_R/d\Delta\eta$, since both are for unidentified charged hadrons, and the empirical verification is based on the data from the same experimental group (PHOBOS).

After the phenomenological relationship between $dN^{ch}_R/d\Delta\eta$ and $dN^{ch}/d\eta$ is established in Sec.\ 2, we give an interpretation of the phenomenon in Sec.\ 3. It is not our objective to give a review of all other models that can reproduce the data on the ridge structure and assess their likelihood to explain the empirical observation made in Sec.\ 2.
We offer only to show the possibility, contrary to conventional wisdom,  that there is no need for long-range longitudinal correlation. Our conclusion is given in Sec.\ 4.

\section{Comparison between Ridge and Inclusive Distributions}

Our focus is on the PHOBOS data on two-particle correlation measured with a trigger particle having transverse momentum $p_T^{\rm trig} > 2.5$ GeV/C in Au + Au collisions at $\sqrt{s_{NN}} = 200$ GeV \cite{ba}.  The pseudorapidity acceptance of the trigger is $0 < \eta^{\rm trig} < 1.5$.  The per-trigger ridge yield integrated over $|\Delta\phi| < 1$, denoted by $(1/N^{\rm trig})dN_R^{ch}/d\Delta\eta$, includes all charged hadrons with $p_T^a\ ^>_\sim\ 7$ MeV/c at $\eta^a = 3$ and $p_T^a\ ^>_\sim\ 35$ MeV/c at $\eta^a = 0$, where the superscript $a$ stands for associated particle in the ridge.  For simplicity we use the notation $\eta^{\rm trig} = \eta_1$, $\eta^a = \eta_2$, $\Delta\eta = \eta_2 - \eta_1$, $\phi^{\rm trig} = \phi_1$, $\phi^a = \phi_2$, $\Delta\phi = \phi_2 - \phi_1$.  Since all ridge particles are included in the range $|\Delta\phi| < 1$, no further consideration is given to the $\phi$ dependence.  Our aim is to relate the ridge distribution in $\Delta\eta$ to the single-particle distribution in $\eta$.  We first make a phenomenological observation using only PHOBOS data for both distributions.  After showing their relationship, we then make an interpretation that does not involve extensive modeling.

To do meaningful comparison, it is important to use single-particle $\eta$ distribution, $dN^{ch}/d\eta$, that has the same kinematical constraints as the ridge distribution.  That is, it involves an  integration over $p_T$ and a sum over all charged hadrons
\begin{eqnarray}
{dN^{ch}\over d\eta} = \sum_h \int dp_Tp_T\rho_1^h (\eta,p_T)  ,    \label{1}
\end{eqnarray}
where $\rho_1^h(\eta,\pt)=dN^h/\pt d\pt d\eta$, and the lower limit of the $p_T$ integration is $35(1-\eta/3.75)$ MeV/c in keeping with the acceptance window of $p_T^a$ \cite{ba}.  The data on $(1/N^{\rm trig})dN_R^{ch}/d\Delta\eta$ are for 0-30\% centrality.  PHOBOS has the appropriate $dN^{ch}/d\eta$ for 0-6\%, 6-15\%, 15-25\% and 25-35\% centralities \cite{bb}, as shown in Fig.\ 1(a). Thus we average them over those four bins.  The result is shown in Fig.\ 1(b) by the small circles for 0-30\% centrality.  Those points are fitted by the three Gaussian \dis s, located at $\eta=0$ and $\pm \hat\eta$,
\begin{eqnarray}
{dN^{ch}\over d\eta} = A \{\exp[-\eta^2/{2\sigma_0^2}] + a_1\exp[-(\eta-\hat\eta)^2/{2\sigma_1^2}] + a_1\exp[-(\eta+\hat\eta)^2/{2\sigma_1^2}]\}  \label{2}
\end{eqnarray}
shown by the solid (red) line in that figure with $A=468, \sigma_0=2.69,  a_1=0.31, \hat\eta = 2.43, \sigma_1=1.15$.  The dashed line shows the central Gaussian, while the dash-dotted line shows the two side Gaussians.
The purpose of the fit is only to give an analytic representation of $dN^{ch}/d\eta$ to be used for comparison with the ridge distribution.  Note that $\hat\eta$ is large $(>2)$, and thus stretches the width of the $\eta$ \dis.

\begin{figure}[tbph]
\vspace*{.5cm}
\includegraphics[width=.9\textwidth,clip]{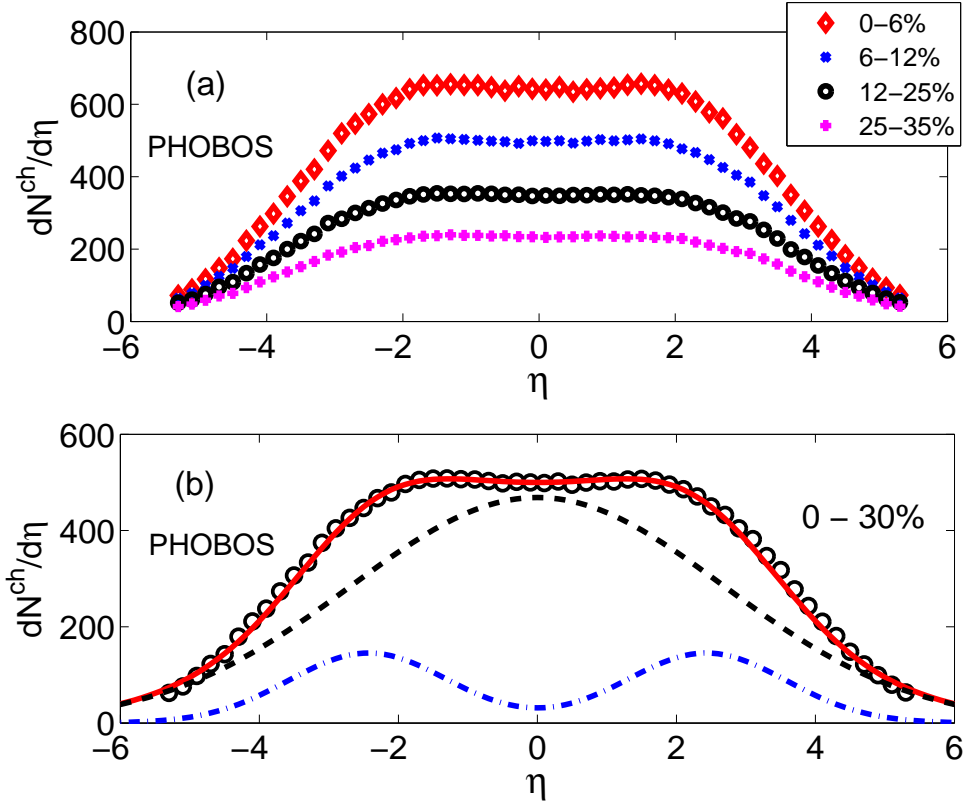}
%\vspace*{-1cm}
\caption{(Color online) Pseudorapidity \dis\ in Au-Au collisions at $\sqrt{s_{NN}}=200$ GeV for (a) various centrality bins  and (b) 0-30\% centrality. Data are from Ref.\ \cite{bb}. The (red) line in (b) is a fit using Eq.\ (\ref{2}), whose first term is represented by the  dashed line and the other two terms by the dash-dotted line.}
\end{figure}

We now propose the formula
\begin{eqnarray}
{1\over N^{\rm trig}} {dN_R^{ch}\over d\Delta\eta} =r\int_0^{1.5} d\eta_1  \left. {dN^{ch}\over d\eta_2} \right|_{\eta_2 = \eta_1 + \Delta\eta},     \label{3}
\end{eqnarray}
where $r$ is a parameter that summarizes all the experimental conditions that lead to the magnitude of the ridge distribution measured relative to the single-particle distribution. In particular, $r$ does not depend on $\eta_1$ or $\eta_2$; otherwise, the equation is meaningless in comparing the $\eta$ dependencies.

There is no theoretical input in Eq.\ (\ref{3}), except for the question behind the proposal:  how much of the $\Delta\eta$ distribution can be accounted for by just a mapping of $dN^{ch}/d\eta_2$ with a shift due to the definition $\Delta\eta = \eta_2 - \eta_1$, and a smearing due to the trigger acceptance, $0 < \eta_1 < 1.5$?  Another way of asking the question is:  if the experimental statistics were high enough so that the trigger's $\eta$ range can be very narrow around $\eta_1 = 0$, how much difference would there be between the ridge distribution in $\Delta\eta$ and the pseudorapidity distribution in $\eta$?

\begin{figure}[tbph]
%\vspace*{-3cm}
\includegraphics[width=1\textwidth,clip]{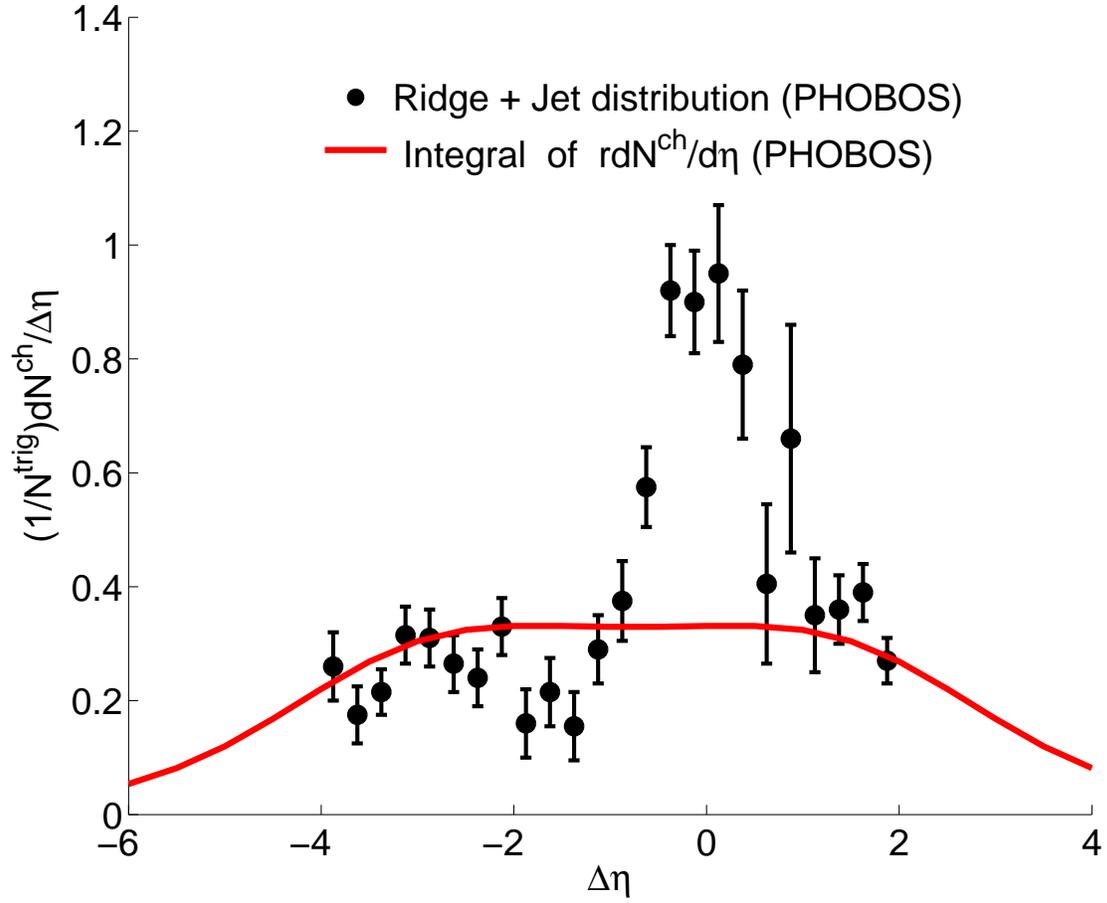}
\vspace*{-1.2cm}
\caption{(Color online) Two-particle correlation of charged particles. Data are from Ref.\ \cite{ba} that include both ridge and jet components. The line is  a plot according to Eq.\ (\ref{3}) using $\eta$ \dis\ from Fig.\ 1 \cite{bb}.}
\end{figure}

The proposed formula in Eq.\ (\ref{3}) is tested by substituting the fit of $dN^{ch}/d\eta$ according to Eq.\ (\ref{2}) into the integrand on the right-hand side.  The result is shown in Fig.\ 2 with $r$ being adjusted to fit the height of the ridge distribution; its value is $4.4\times 10^{-4}$. The peak in the data around $\Delta\eta=0$ is, of course, due to the jet component associated with the trigger jet and is not relevant to our comparison here.  It is evident that the large $\Delta\eta$ distribution is well reproduced by Eq.\ (\ref{3}).  In qualitative terms the width of the ridge distribution is due partly to the width of $dN^{ch}/d\eta$ and partly to the smearing of $\eta_1$, which adds another 1.5 to the width.  No intrinsic dynamics of long-range longitudinal correlation has been put in.  Note that the center of the plateau in $\Delta\eta$ is at $-0.75$, which is the average of the shift due to $\eta_1$ being integrated from 0 to 1.5. It suggests that if $\eta_1$ were fixed at $\eta_1\approx 0$ when abundant data become available, then the width of $dN_R^{ch}/d\Delta\eta$ would be only as wide as that of the single-particle $dN^{ch}/d\eta$. No theoretical prejudice has influence these observations.

\section{Interpretation of Phenomenological Observation}

	We now consider an interpretation of what Eq. (\ref{3}) implies, given the empirical support for its validity from Fig.\ 2. Since the observed ridge \dis\ integrates over trigger $\eta$, we write it as
\bq
{1\over N^{\rm trig}}{dN_R^{ch}\over d\Delta\eta}=\int_0^{1.5} d\eta_1\sum_{h_2} \int dp_2 p_2\left. R^{h_2}(\eta_1,\eta_2,p_2) \right|_{\eta_2 = \eta_1 + \Delta\eta},     \label{4}
\eq
where we exhibit also explicitly the sum over the hadron type of the ridge particle $h_2$ and the integral over its momentum, denoted by $p_2$. According to the definition of correlation $C_2(1,2)=\rho_2(1,2)-\rho_1(1)\rho_1(2)$, we can express the per-trigger ridge correlation as
\bq
R^{h_2}(\eta_1,\eta_2,p_2) = \sum_{h_1} \int dp_1 p_1 {\rho_2^{h_1h_2{(B+R)}}(\eta_1,p_1,\eta_2,p_2)\over \rho_1^{h_1}(\eta_1,p_1)} -\rho_1^{h_2{(B)}}(\eta_2,p_2),  \label{5}
\eq
where $B$ and $R$ in the superscript denote background and ridge, respectively. The jet component
in the associated-particle \dis\ is excluded in Eq.\ (\ref{5}).

On the other hand, with Eq.\ (\ref{1}) substituted into Eq.\ (\ref{3}) we  have, using $\eta_2$ and $p_2$ instead of 
$\eta$ and $p_T$,
\bq
{1\over N^{\rm trig}}{dN_R^{ch}\over d\Delta\eta}=\int_0^{1.5} d\eta_1\sum_{h_2} \int dp_2 p_2\left. r \rho_1^{h_2}(\eta_2,p_2) \right|_{\eta_2 = \eta_1 + \Delta\eta}.     \label{6}
\eq
Comparing Eq.\ (\ref{6}) to (\ref{4}) we see that the ridge \dis\ $R^{h_2}(\eta_1,\eta_2,p_2)$ is to be related to the phenomenological quantity $r\rho_1^{h_2}(\eta_2,p_2)$. Thus the crux of the relationship between the ridge and inclusive \dis s involves the interpretation of $r\rho_1^{h_2}$. To that end let us first write $\rho_1^{h_2}$ in the form
\bq
\rho_1^{h_2}(\eta_2,p_2)={dN^{h_2}\over d\eta_2p_2dp_2}=H^{h_2}(\eta_2,p_2) V(p_2) ,  \label{7}
\eq
where $V(p_2)=e^{-p_2/T}$ is the transverse component that contains the explicit exponential behavior of $p_2$. Although $H^{h_2}(\eta_2,p_2)$ has some mild $p_2$ dependence due mainly to mass effects of $h_2$, the average transverse momentum $\left<p_2\right>$ is determined primarily by the inverse slope $T$ and is not dependent on $\eta_2$. This is an approximate statement that is based on the BRAHMS data \cite{bea}, which show that $\left<p_T\right>$ is essentially independent of rapidity. Since $r$ serves as the phenomenological bridge between $R^{h_2}$ and $\rho_1^{h_2}$, the key question to address is: which of the two components, the longitudinal $H^{h_2}(\eta_2,p_2)$ or the transverse $V(p_2)$, does the two-particle correlation exert its most important influence in relating $R^{h_2}$ to $\rho_1^{h_2}$?

If there is longitudinal correlation from early times as in \cite{ad,gmm,10}, then its effect must be to convert $H^{h_2}(\eta_2,p_2)$ to $R^{h_2}(\eta_1,\eta_2,p_2)$. In that case $V(p_2)$ is relegated to the secondary role due to radial flow (which is, nevertheless, essential in explaining the $\Delta\phi$ restriction as in Refs.\ \cite{gmm,sv,shu,12,17}). On the other hand, if there is no intrinsic long-range longitudinal correlation, then $H^{h_2}(\eta_2,p_2)$ is unaffected, and the ridge can only arise from the change in the transverse component, $V(p_2)$. Without phenomenology one would think that the first option is more reasonable, when $|\Delta\eta|\sim 4$ is regarded as large, and especially when there is an inclination based on theoretical ideas that there is long-range correlation. With the ridge phenomenology encapsuled in Eq.\ (\ref{3}) pointing to direct relevance of $H^{h_2}(\eta_2,p_2)$, the question becomes that of asking:   $|\Delta\eta|$ is large compared to what? If it is now recognized that $|\Delta\eta|$ is not large compared to the $\eta$ range of $\rho_1^{h_2}(\eta_2,p_2)$ after the widening due to $\eta_1$ smearing (remarked at the end of the previous section) is taken into account,
then the need for a long-range dynamical correlation to account for the structure of $R^{h_2}(\eta_1,\eta_2,p_2)$ is  lost. We describe below a possible explanation based on the second option of no long-range correlation. The key is to accept the suggestion of the data that the unmodified  longitudinal component $H^{h_2}(\eta_2,p_2)$ is sufficient.

A series of articles have treated the subject of ridge formation in the recombination model \cite{hy}, beginning with (a) the early observation of pedestal in jet correlation \cite{12,11}, to (b) its effects on azimuthal anisotropy of single-particle \dis\ at mid-rapidity \cite{13,14}, and then to (c) the dependence on the azimuthal angle $\phi_s$ of the trigger relative to the reaction plane \cite{15,16,17,18}. Forward productions in d-Au and Au-Au collisions have also been studied in \cite{19,20}. Our consideration here of ridge formation at $|\Delta\eta|>2$ is an extension of earlier studies with the common theme that ridges are formed as a consequence of energy loss by semihard or hard partons as they traverse the medium. The details involve careful treatment of the hadronization process with attention given to both the longitudinal and transverse components. The $\phi$ dependence has been studied thoroughly in \cite{17,18}, and the $\eta$ dependence should take into account of
 the experimental fact that the $p/\pi$ ratio can be large $(> 2.5)$ at large $\eta$ \cite{23} so that $H^{h_2}(\eta_2,p_2)$  in Eq.\ (\ref{7}) can be properly reproduced, which is a subject to be reported elsewhere \cite{21}. 
 
 For our purpose here we need not repeat all the details, but state only that the effect of jets on the medium is that the energy loss enhances the thermal fluctuations of the soft partons in the early stage (before Hubble-like expansion takes place) so that even at $|\Delta\eta|>2$ the transverse component $V(p_2)$ in Eq.\ (\ref{7}) is sensitive to the thermal enhancement. The implementation of that basic physical idea on the formation of ridge particles (called pedestal at the time) was first carried out in Ref.\ \cite{12} in terms of thermal-shower recombination for the trigger, and of thermal-thermal recombination for the associated particles. The result of the collective investigation \cite{12,13,14,18} can be summarized by
\bq
R^{h_2}(\eta_1,\eta_2,p_2)=c H^{h_2}(\eta_2,p_2) V_R(\eta_1,\eta_2,p_2),  \label{8}
\eq
where
\bq
 V_R(\eta_1,\eta_2,p_2)= V_{B+R}(\eta_1,\eta_2,p_2)- V_B(p_2),  \label{9}
 \eq
 which has the structure of Eq.\ (\ref{5}), but only in the transverse part. 
The constant $c$  characterizes the magnitude of the ridge, which can depend  on many factors that  include the fluctuations in the initial configuration, the details of correlation dynamics, the experimental cuts, the
$\Delta\phi$ interval where the ridge is formed and the related scheme of background subtraction. Its value (that was not calculated) does not affect the relationship between the $\eta$ dependencies of the two sides of Eq.\ (\ref{8}).
Of more relevant concern to us are the transverse components, which may be written more explicitly as
 \bq
 V_B(p_2)=\exp(-p_2/T),  \qquad  V_{B+R}(\eta_1,\eta_2,p_2)=\exp[-p_2/T'(\Delta\eta)].  \label{10}
 \eq
 Thus $T$ is the inverse slope of the inclusive thermal \dis, regarded in Eq.\ (\ref{9}) as the background, while $T'>T$ represents the enhancement effect due to the semihard or hard parton. The parton's trajectory toward the near-side surface has a conical vicinity in which the medium has increased thermal activity because of the energy loss of the parton.
 In the initial configuration of  spatial uncertainty in $\Delta z$ we allow  the right-moving soft partons to start from the left side of the thermally enhanced cone, and similarly the left-moving soft partons to start from the right side, so that the transverse \dis\ of the soft partons can change from $T$ to $T'$ due to their passage through the enhanced cone. This is transverse broadening as in the conventional interpretation of the Cronin effect \cite{22} but only for associated particles on the near side within a restricted region of $\Delta\phi$ around the trigger $\phi_1$.
 Furthermore,  the polar-angular relationship between the soft partons and the semihard parton can fluctuate significantly during the broadening process that affect both the longitudinal and transverse momenta of the associated particles.
 $T'$ is a measure of the average effect on those particles. It should not depend significantly on $\Delta\eta$, just as the Cronin effect does not have sensitive dependence on the longitudinal momentum. Thus we shall set $T'(\Delta\eta)$ to be roughly constant in $\Delta\eta$ to about the same degree of approximation as $T$  in $V(p_2)$ is independent of $\eta_2$ in Eq.\ (\ref{7}). To let $T'$ be independent of $\Delta\eta$ is not equivalent to putting in by hand long-range longitudinal correlation; that is analogous to the proposition in the flux-tube model that assuming similar radial flow for all parts of a tube of partons (so that they all gain the same average $\pt$) does not imply that long-range longitudinal correlation is inserted by hand. There is, however, transverse correlation in the sense that the $p_2$ dependence of $V_R(p_2)$ is the same at all $\eta_2$. That is  similar to the familiar phenomenon that rising tide raises all boats, where the boat heights are transverse to the spatial separation that is longitudinal,  the former being correlated, the latter not. Experimentally, for trigger momentum in the interval $4<p_1<6$ GeV/c, it is found that $\Delta T\equiv T'-T$ is less than 20\% of $T$ \cite{bia}.

Putting the various parts together, we have 
\bq
V_R(p_2) = e^{-p_2/T}(e^{p_2/T''}-1) ,  \qquad\quad  {1\over T''}={1\over T}-{1\over T'}={\Delta T\over TT'}.   \label{11}
\eq
This expression was first obtained in Refs.\ \cite{13,14} as a description of the ridge \dis\ without trigger. It was noted there that $V_R(p_T)\to 0$ as $p_T\to 0$, and that $p_T/T''$ sets the scale for $v_2(p_T,b)$ for $p_T<0.5$ GeV/c in agreement with the data on elliptic flow. Since the lower limit of the acceptance window for associated-particle momentum in the PHOBOS experiment is low \cite{ba}, the relevant values of $T$ and $T'$ may not be the same as those measured by STAR \cite{bia}. For the purpose of quantitative orientation they may be  set  at $T=0.28$ GeV and $T'=0.32$ GeV, so that $T''=2.24$ GeV. The exact values are not important to our qualitative conclusion to be drawn below.

We now substitute Eq.\ (\ref{8}) in (\ref{4}) and use (\ref{7}) with $V(p_2)$ identified as $V_B(p_2)$ to obtain
\bq
{1\over N^{\rm trig}}{dN_R^{ch}\over d\Delta\eta}=\int_0^{1.5} d\eta_1\sum_{h_2} \int dp_2 p_2\left. {cV_R(p_2)\over V_B(p_2)} \rho_1^{h_2}(\eta_2,p_2) \right|_{\eta_2 = \eta_1 + \Delta\eta}.     \label{12}
\eq
Comparing this equation with Eq.\ (\ref{6}), we come to the conclusion that $r$ is a phenomenological approximation of  $c V_R(p_2) / V_B(p_2)$ in the region where it contributes most to the integral over $p_2$. By itself $ V_R(p_2) / V_B(p_2)$ increases with $p_2$ as $e^{p_2/T''}-1$, but it is damped more severely by the exponential decrease of $\rho_1^{h_2}(\eta_2,p_2)$ for $p_2>1$ GeV/c, since $T''\gg T$. Thus $c V_R(p_2) / V_B(p_2)$ may be approximated by a constant $r$ in the region where the integrand is maximum at around $p_2\sim 0.5$ GeV/c. In so doing, we obtain Eq.\ (\ref{6}) and therefore the phenomenological relation given by Eq.\ (\ref{3}).

Equation (\ref{12}) implies that there is transverse correlation, but no explicit longitudinal correlation beyond what is implicitly contained in $\rho_1^{h_2}$. The transverse correlation that we refer to is not what one usually associates with the correlation between hadrons in the fragments of a high-$\pt$ jet. All of those fragments are in a small range of $\Delta\eta$ and have transverse-momentum fractions that are correlated. They populate the peak in Fig.\ 2. In our problem about the ridge we have been concerned with the transverse momentum of a particle associated with a trigger outside that peak. The former reveals the effect of the medium on the  jet, while the latter reveals the effect of the jet on the medium. That is the basic difference between the jet and ridge components of the associated particles. Since semihard or hard scattering takes place early, transverse broadening can take place for soft partons (the medium) moving through the interaction zone, leading to the ridge structure.

\section{Conclusion}

An issue that this study has brought up is the usage of the word ``large'' in referring to the  range of $\Delta\eta$ in the ridge structure. Our  phenomenological observation in Eq.\ (\ref{3}), substantiated by Figs.\ 1 and 2, does not reveal any quantitative definition of what large $\Delta\eta$ means. To say that $|\Delta\eta|>2$ is large is a figurative description until it is followed by a statement on dynamics.
 To be able to relate large $\Delta\eta$ to dynamical long-range correlation is a worthy theoretical endeavor, but more can be added to its phenomenological relevance if it can also elucidate the empirical connection between the two sides of Eq.\ (\ref{3}).

Since the pseudorapidity \dis\ $dN^{ch}/d\eta$ involves an integration over $p_T$ and sum over charged hadrons, as expressed in Eq.\ (\ref{1}), a full understanding of its $\eta$ dependence must take into account the experimental fact from BRAHMS that protons are produced more than twice as many as  pions at $\eta> 2$ and $\pt\ ^>_\sim\ 2$ GeV/c in central collisions \cite{23}. In fact, it is seen in Fig.\ 1(a) that the PHOBOS data on $dN^{ch}/d\eta$ at 0-6\% centrality show a bump at $\eta\approx 2$, which is undoubtedly related to proton production. The widening of the $\eta$ \dis\ is a hadronization problem that involves late-time physics. Long-range longitudinal correlation is early-time physics. To connect the two is a theoretical problem that remains to be done. Most  models emphasizes one or the other. Equation (\ref{3}) that connects $dN_R^{ch}/d\Delta\eta$ to $dN^{ch}/d\eta$ is an empirical statement without any direct revelation  on the nature of the dynamics throughout all times of the collision process. It is therefore a worthy goal to explain that connection from all approaches. 

The approach that we have taken provides a possible solution involving no long-range longitudinal correlation in addition to what accounts for the single-particle \dis\ $\rho_1$.
Since $\rho_1$ is an integral over the two-particle \dis\ $\rho_2$, whatever intrinsic correlation that exists among the constituents must contribute to the properties of $\rho_1$. In this paper we have not delved into the details of $\rho_1$; we have only made the observation that, given $\rho_1$, it is possible to get $dN_R^{ch}/d\Delta\eta$ without additional longitudinal correlation of any range. It gives a hint on how the ridge structure is to be understood. But as with any model on particle production, the burden of proof is then shifted heavily toward an explanation of $\rho_1$ in all its details, only the integrated form of which is shown in Fig.\ 1.

Although we have given a simple interpretation of the phenomenological quantity $r$, the importance of this work leans more toward the finding of the relationship between $dN_R^{ch}/d\Delta\eta$ and $dN^{ch}/d\eta$ and of the substantiation it receives from the two pieces of data from PHOBOS exhibited in Figs.\ 1 and 2.  An extension of our interpretation would lead naturally to the prediction that similar phenomenon will be found at LHC.

\section*{Acknowledgment}
This work was supported, in part,  by the U.\ S.\  Department of Energy under Grant No. DE-FG02-92ER40972.

%\newpage


\begin{thebibliography}{99}

%1
\bibitem{ja}
J.\ Adams {\it et al.}, (STAR Collaboration), Phys.\ Rev.\ C {\bf 73}, 064907 (2006).

%2
\bibitem{bia}
B.\ I.\ Abelev {\it et al.}, (STAR Collaboration), Phys.\ Rev.\ C {\bf 80}, 064912 (2009).

%3
\bibitem{ba}
B.\ Aver {\it et al.}, (PHOBOS Collaboration), Phys.\ Rev.\ Lett.\ {\bf 104}, 062301 (2010).

%4
\bibitem{bia1}
B.\ I.\ Abelev {\it et al.}, (STAR Collaboration), Phys.\ Rev.\ Lett.\  {\bf 105}, 022301 (2010).

%5
\bibitem{md}
M.\ Daugherity (for STAR Collaboration), J.\ Phys.\ G: Nucl.\ Part.\ Phys.\ {\bf 35}, 104090 (2008).

%6
\bibitem{cms}
CMS Collaboration, J.\ High Energy Physics (to be published), arXiv: 1009.4122.

%7
\bibitem{ad}
A.\ Dumitru, F.\ Gelis, L.\ McLerran, and R.\ Venugopalan, Nucl.\ Phys.\ {\bf A810}, 91 (2008).

%8
\bibitem{gmm}
S.\ Gavin, L.\ McLerran and G.\ Moschelli, Phys.\ Rev.\ C {\bf 79}, 051902(R) (2009);
G.\ Moschelli and S.\ Gavin, Nucl.\ Phys.\ A {\bf 836}, 43 (2010).

%9
\bibitem{bb}
B.\ B.\ Back {\it et al.} (PHOBOS Collaboration), Phys.\ Rev.\ Lett.\ {\bf 91}, 052303 (2003).

%10
 \bibitem{bea}
  I.\ G.\ Bearden {\it et al} (BRAHMS Collaboration) Phys.\ Rev.\ Lett. {\bf 93}, 102301 (2004); {\bf 
94}, 162301 (2005).

%11
\bibitem{10}
K.\ Dusling, F.\ Gelis, T.\ Lappi and R.\ Venugopalan, arXiv:  0911.2720

%12
\bibitem{sv}
S.\ A.\ Voloshin, Phys.\ Lett.\ B {\bf 632}, 490 (2006).

%13
\bibitem{shu}
E.\ Shuryak,  Phys.\ Rev.\ C {\bf 76}, 047901 (2007).


%14
\bibitem{12}
C.\ B.\ Chiu and R.\ C.\ Hwa, Phys.\ Rev.\ C {\bf 72}, 034903 (2005).

%15
\bibitem{17}
C.\ B.\ Chiu and R.\ C.\ Hwa, Phys.\ Rev.\ C {\bf 79}, 034901 (2009).

%16
\bibitem{hy}
R.\ C.\ Hwa and C.\ B.\ Yang, Phys.\ Rev.\ C {\bf 70}, 024905 (2004).


%17
\bibitem{11}
J.\ Adams {\it et al.} (STAR Collaboration), Phys.\ Rev.\ Lett.\ {\bf 
95}, 152301 (2005).



%18
\bibitem{13} 
R.\ C.\ Hwa, Phys.\ Lett.\ B {\bf 666}, 228 (2008).

%19
\bibitem{14}
C.\ B.\ Chiu, R.\ C.\ Hwa, and C.\ B.\ Yang, Phys.\ Rev.\ C {\bf 78}, 044903 (2008).

%20
\bibitem{15}
A.\ Feng,  (for STAR Collaboration), talk given at Quark Matter 2008, Jaipur, India (2008), J.\ Phys.\ G: Nucl.\ Part.\ Phys.\ {\bf 35}, 104082 (2008), arXiv: 0807.4606.

%21
\bibitem{16}
H.\ Agakishiev {et al.} (STAR Collaboration), arXiv: 1010.0690.


%22
\bibitem{18}
R.\ C.\ Hwa and L.\ Zhu, Phys.\ Rev.\ C {\bf 81}, 034904 (2010).

%23
\bibitem{19}
R.\ C.\ Hwa, C.\ B.\ Yang, and R.\ J.\ Fries, Phys.\ Rev.\ C {\bf 71}, 024902 (2005).

%24
\bibitem{20}
R.\ C.\ Hwa and L.\ Zhu, Phys.\ Rev.\ C {\bf 78}, 024907 (2008).

%25
\bibitem{23}
I.\ G.\ Arsene {\it et al.} (BRAHMS  Collaboration), Phys.\ Lett.\ B {\bf 684}, 22 (2010).

%26
\bibitem{21}
C.\ B.\ Chiu and R.\ C.\ Hwa, (to appear).



%22
\bibitem{22}
J.\ W.\ Cronin {\it et al.}, Phys.\ Rev.\ D {\bf 11}, 3105 (1975).

\end{thebibliography}
\end{document}